\documentclass[twoside]{article}
\usepackage{qic,epsfig}
\usepackage{theorem}
\usepackage{enumerate}
\usepackage{amsmath}
\usepackage{amssymb}

\newcommand{\lv}{\left \vert}
\newcommand{\rv}{\right \vert}
\newcommand{\la}{\left \langle}
\newcommand{\ra}{\right \rangle}
\newcommand{\ket}[1]{\lv #1 \ra}
\newcommand{\bra}[1]{\la #1 \rv}

\newcommand{\B}{\mathfrak{B}}
\newcommand{\Hi}{\mathcal{H}}

\newcommand{\Tr}{\mathrm{tr}}
\theorembodyfont{\rmfamily}
\newtheorem{Theorem}{\textit{Theorem} }

\newtheorem{Definition}{\textit{Definition} }
\newtheorem{Corollary}{\textit{Corollary}}
\newtheorem{Lemma}{\textit{Lemma}}
\newtheorem{Proof}{\textit{Proof}}

\begin{document}
\setlength{\textheight}{8.0truein}    

\runninghead{$\epsilon$-convertibility  of entangled states and
extension of Schmidt rank in infinite-dimensional systems}
            {M. Owari, S. L. Braunstein, K. Nemoto, M. Murao}

\normalsize\textlineskip
\thispagestyle{empty}
\setcounter{page}{1}


\vspace*{0.88truein}

\alphfootnote

\fpage{1}

\centerline{\bf {\Large $\epsilon$}-CONVERTIBILITY OF ENTANGLED STATES AND EXTENSION}
\vspace*{0.035truein} \centerline{\bf
   OF SCHMIDT
  RANK IN INFINITE-DIMENSIONAL SYSTEMS} \vspace*{0.37truein}
\centerline{\footnotesize Masaki Owari} \vspace*{0.015truein}
\centerline{\footnotesize\it Collaborative Institute for Nano Quantum Information Electronics, The University
of Tokyo\footnote{The major part of this work was done when MO was in Department of Physics, Graduate School of Science, The University of Tokyo.}} 
\baselineskip=10pt \centerline{\footnotesize\it  Tokyo
113-0033, Japan} \vspace*{10pt} \centerline{\footnotesize Samuel
L. Braunstein} \vspace*{0.015truein} \centerline{\footnotesize\it
Computer Science, University of York} \baselineskip=10pt
\centerline{\footnotesize\it York YO10 5DD, United Kingdom}
\vspace*{10pt} \centerline{\footnotesize Kae Nemoto}
\vspace*{0.015truein} \centerline{\footnotesize\it National
Institute of Informatics } \baselineskip=10pt
\centerline{\footnotesize\it  Tokyo 101-8430, Japan}
\vspace*{10pt} \centerline{\footnotesize Mio Murao}
\vspace*{0.015truein} \centerline{\footnotesize\it Department of
Physics, The University of Tokyo } \baselineskip=10pt
\centerline{\footnotesize\it  Tokyo 113-0033, Japan}
\centerline{\footnotesize\it PRESTO, JST } \baselineskip=10pt
\centerline{\footnotesize\it  Kawaguchi, Saitama 332-0012, Japan}

\vspace*{0.225truein}

\vspace*{0.21truein}

\abstracts{By introducing the concept of
$\epsilon$-convertibility, we extend Nielsen's and Vidal's
theorems to the entanglement transformation of
infinite-dimensional systems. Using an infinite-dimensional
version of Vidal's theorem we derive a new stochastic-LOCC (SLOCC)
monotone which can be considered as an extension of the Schmidt
rank. We show that states with polynomially-damped Schmidt
coefficients belong to a higher rank of entanglement class in
terms of SLOCC convertibility. For the case of Hilbert spaces of
countable, but infinite dimensionality, we show that there are
actually an uncountable number of classes of pure
non-interconvertible bipartite entangled states. }{}{}

\vspace*{10pt}

 \vspace*{3pt}

\vspace*{1pt}\textlineskip    

\section{Introduction}
%
Entanglement is one of the central topics in quantum information, and
has both physical and information scientific aspects. In
particular, entanglement involves quantum non-local correlations
which have been of interest in physics \cite{epr}, and also acts
as a resource of quantum information processing for informatics
\cite{resource, quantum communication}. Thus, the characterization
of the entanglement of physical systems is important from both a
physical and information scientific viewpoint.

Mathematically, physical systems can be categorized into two
classes, that is, finite-dimensional systems which can be treated
in the framework of conventional linear algebra, and
infinite-dimensional systems which need to be treated in the
framework of functional analysis \cite{neumann, functional
analysis}. It is therefore worthwhile to know whether such a
mathematical difference of systems can make an essential
difference in the properties of entanglement in these systems.
Indeed, this may provide an answer to the question of what is the
essential difference between the physics of finite-dimensional
systems and the physics of infinite-dimensional systems from the
viewpoint of non-local correlations. Moreover, from the
information-theoretic viewpoint, if such a difference exists,
there may be the possibility that we can achieve an information
processing in infinite-dimensional systems which cannot be
achieved in finite-dimensional systems.

In this paper, we mainly focus on seeking a difference between the
properties of entanglement of finite-dimensional systems and those of
infinite-dimensional systems. Since much work on the
characterization of bipartite entangled states has been done in
finite-dimensional systems \cite{bennett, majorization, vidal,
monotone}, we concentrate our efforts on the characterization of
bipartite entangled states in infinite-dimensional systems, and
try to find a difference in the properties of their entanglement.

So far, research on the characterization of entanglement in
infinite dimensions has been done in the form of separability
criteria \cite{separability}, Gaussian LOCC convertibility
\cite{gaussian, gaussian distill}, and entanglement measures
\cite{entanglement measure}. The separability criteria gives us a
way of judging whether or not a given state is entangled. The
Gaussian LOCC convertibility gives the detailed structure of the
strength of entanglement of Gaussian states. Entanglement measures
give an approximated strength of entanglement in the limit of an
asymptotic infinite number of copies.

The above research mainly is concerned with Gaussian states and
Gaussian operations, and unique properties of infinite-dimensional
entanglement do not appear clearly in this regime. Therefore, in
order to find a property unique to infinite-dimensional
entanglement it is important to investigate the strength of
entanglement more precisely for a broader class of states and
operations.

The strength of entanglement is defined by means of the
convertibility between entangled states under local operations,
\textit{e.g.}, local operations and classical communication
(LOCC), stochastic-LOCC (SLOCC), or the positive partial transpose
(PPT) operation \cite{LOCC,uniqueness}. Among such local
operations we mainly focus in this paper on SLOCC, and investigate
the SLOCC convertibility of entangled states in
infinite-dimensional systems without any assumption for states or
operations to find a unique property of entanglement in
infinite-dimensional systems.

When we consider SLOCC convertibility for infinite-dimensional
systems, there are at least two difficulties, namely, the problem
of continuity and the problem of a potentially infinite cost for
classical communication. In order to avoid such difficulties, we
propose a new definition of state convertibility that we call
$\epsilon$-convertibility. We define $\epsilon$-convertibility as
the convertibility of states in an approximated setting by means
of the trace norm.  Then, within the framework of
$\epsilon$-convertibility, we investigate SLOCC convertibility in
infinite-dimensional systems and show a fundamental difference of
SLOCC convertibility between infinite and finite-dimensional
systems.

The paper is organized as follows: in section \ref{infinite
Nielsen}, we define the $\epsilon$-convertibility of LOCC and
SLOCC, and show how to avoid the problems of discontinuity and
infinitely-costly classical communication. Then within the
framework of $\epsilon$-convertibility, we give the
infinite-dimensional extensions and proofs of Nielsen's and
Vidal's theorem, which give the necessary and sufficient
conditions of LOCC and SLOCC convertibility, respectively. In
section \ref{Extension}, first we define monotones (monotonic
functions) of SLOCC convertibility, which can be considered an
extension of the Schmidt rank for infinite-dimensional systems,
then by means of this monotone, we investigate the SLOCC
convertibility for infinite-dimensional systems. We show that the
cardinal number of the quotient set of states by SLOCC
convertibility is greater than or equal to the cardinal number of
the continuum, and also show that however many (finite) copies of
exponentially-damped states (states with exponentially damped
Schmidt coefficients) there are, they cannot be converted into
even a single copy of a polynomially-damped state (a state with
polynomially-damped Schmidt coefficients). Such properties do not
exist in finite-dimensional systems and are actually unique to
infinite-dimensional systems.


\section{$\epsilon$-convertibility} \label{infinite Nielsen}

In this paper we consider the bipartite infinite-dimensional
system $\Hi = \Hi _A \otimes \Hi _B$ where $\dim \Hi _A = \dim \Hi
_B = \infty$ and we shall assume that $\Hi _A$ and $\Hi _B$ are
separable. By $\B (\Hi)$ we denote the Banach space of all bounded
operators on $\Hi$. If we use the term LOCC, we will always assume
that operations succeed \textit{with unit probability}
\cite{uniqueness}. On the other hand we use the term SLOCC in the
case where operations work with a finite probability less than
unity. For simplicity we use at most countably infinite POVMs as
the element of an LOCC (or SLOCC), $\{A_i\}_{i=1}^{\infty}$,
$A_i\in\B(\Hi)$, $\sum_{i\in \mathbb{N}}A ^{\dagger}A =
(\textbf{or} \leq )I$ (corresponding to ultra-weak convergence).


\subsection{$\epsilon$-convertibility for LOCC and SLOCC}\label{epsilon}

As mentioned in the introduction to give a detailed discussion of
SLOCC convertibility in infinite-dimensional systems, there are at
least two difficulties, namely, discontinuity and infinite
classical communication costs. In this subsection we define
$\epsilon$-convertibility and see that it allows us to avoid the
difficulty of discontinuity. The other difficulty is addressed in
the following subsection.

In infinite-dimensional systems we cannot deny the possibility
that $\ket{\Psi}$ is SLOCC convertible to any neighborhood of
$\ket{\Phi}$  (in terms of strong, or weak topology), but not to
$\ket{\Phi}$ itself. To avoid such a discontinuity, when
considering convertibility among genuinely infinite-dimensional
states (i.e., states with infinitely many non-zero Schmidt
coefficients), we shall identify these neighborhoods with the
state itself. To achieve this we shall extend the definition of
LOCC and SLOCC convertibility to satisfy the above requirement.
Mathematically, we redefine LOCC convertibility as follows:
$\ket{\Psi}$ can be converted to $\ket{\Phi}$ by LOCC, if and only
if for any neighborhood of $\ket{\Phi}$, there exists an LOCC
operation by which $\ket{\Psi}$ is transformed to a state in the
neighborhood of $\ket{\Phi}$. We call this new definition of
convertibility $\epsilon$-convertibility. Below we rigorously
define $\epsilon$-convertibility for LOCC, then we show that this
definition recovers the continuity of convertible probability at
least with some suitable weak meaning.

Before we give the definition of $\epsilon$-convertibility, we
need to choose a topology of the convergence which we use in our
definition. Actually, it is well known that there are many
different topologies defined by associated norms in
infinite-dimensional systems. Therefore, we need to take care to
choose our `distance'. Since we introduced
$\epsilon$-convertibility because of the fundamental impossibility
for discriminating a state $\ket{\Phi}$ from states within
infinitely small neighborhoods of $\ket{\Phi}$, the distance we
consider needs to echo this difficulty with discrimination. We can
easily see that the trace norm possesses such a property as
follows.  Suppose $M$ is an arbitrary POVM element and $\lim
_{n\rightarrow\infty}\|\rho-\rho_n\|_{\rm tr}=0$. Then, $\lim _{n
\rightarrow \infty} | \Tr\; \rho M - \Tr\; \rho _n M| \le \lim _{n
\rightarrow \infty} \| \rho - \rho _n \| _{\rm tr}\|M\|_{\rm op} =
0$, where $\| \cdot \| _{\rm op}$  is the operator norm. Thus, for
all measurements the resulting probability distributions for
$\rho_n$ converge to the resulting probability distribution for
$\rho$. That is, if $\rho_n$ converges $\rho$ in the trace norm,
there is no way to discriminate $\rho$ from $\rho _n$ for
sufficiently large $n$. But this is just the property required of
the distance needed to deal with the discontinuity difficulty in
the definition of $\epsilon$-convertibility. Therefore, we shall
use the trace norm as our distance measure for
$\epsilon$-convertibility.

Following this discussion we rigorously define
$\epsilon$-convertibility for LOCC as:
\begin{Definition}
We say that $\ket{\Psi}$ is $\epsilon$-convertible to $\ket{\Phi}$
by LOCC, if for any $\epsilon > 0$, there exists an LOCC operation
$\Lambda$ which satisfies the condition
$\|\Lambda(\ket{\Psi}\bra{\Psi})-\ket{\Phi}\bra{\Phi}\|_{\rm
tr}<\epsilon$ where $\|\cdot\|_{\rm tr}$ is the trace norm.
\end{Definition}
Similarly, we define $\epsilon$-convertibility for SLOCC as:
\begin{Definition}
\label{epsilon SLOCC} We say that $\ket{\Psi}$ is
$\epsilon$-convertible to $\ket{\Phi}$ by SLOCC with probability
$p > 0$ if for any $\epsilon
>0$, there exists an SLOCC operation $\Lambda $ which satisfies the following
condition, $\| \Lambda ( \ket{\Psi} \bra{\Psi} )/\Tr\; \Lambda (
\ket{\Psi} \bra{\Psi} ) - \ket{\Phi} \bra{\Phi} \| _{\rm tr} <
\epsilon$ and $\Tr\; \Lambda ( \ket{\Psi} \bra{\Psi} ) \ge p$.
\end{Definition}
This definition of $\epsilon$-convertibility under SLOCC means
that {\itshape with more than some fixed non-zero probability
$p$}, $\ket{\Psi}$ can be converted to any neighborhood of
$\ket{\Phi}$ by SLOCC.

Here, we prove by means of $\epsilon$-convertibility that we can
recover enough continuity to achieve a classification of states by
SLOCC convertibility. In infinite-dimensional systems when we
consider SLOCC convertibility it might happen that $\ket{\Psi}$
cannot be converted to $\ket{\Phi}$ and yet there exists a
sequence of $\ket{\Phi _n}$ such that $\ket{\Psi}$ can be
converted to $\ket{\Phi _n}$ with probability $p_n$ and $\lim _{n
\rightarrow \infty} p_n > 0$; however, we cannot discriminate
$\ket{\Phi }$ and $\ket{\Phi _n}$ for large $n$. If this were to
happen then it would be nonsense that $\ket{\Psi}$ could not
converted to $\ket{\Phi}$ by SLOCC.\fnm{{\it a}}\fnt{}{\ To avoid confusion, we add a remark.
We do not know an explicit example of this discontinuity.
However, in infinite dimensional systems, 
it is not trivial whether such discontinuity occurs or not.
Hence, it is better to introduce a modified definition of convertibility under which
we can trivially avoid the discontinuity.}
  However, by means of our new
definition of convertibility, we avoid such a discontinuity. That
is, we can easily show the following continuity property of
$\epsilon$-convertibility.
\begin{Lemma}\label{epsilon SLOCC lemma}
If $\ket{\Psi}$ is not $\epsilon$-convertible to $\ket{\Phi}$ by
SLOCC, but $\ket{\Psi}$ is $\epsilon$-convertible to $\ket{\Phi
_n}$ by SLOCC with probability $p_n$ for all $n$, where $\lim _{n
\rightarrow \infty} \ket{\Phi _n} \bra{\Phi _n} = \ket{\Phi}
\bra{\Phi}$ by the trace norm, then $\lim _{n \rightarrow \infty}
p_n = 0$.
\end{Lemma}
\begin{Proof}
We prove this lemma by contradiction. We assume the following
condition; $\ket{\Psi}$ is not $\epsilon$-convertible to $\ket{\Phi}$,
$\ket{\Psi}$ is $\epsilon$-convertible to $\ket{\Phi _n}$ with
probability $p_n > 0$, where $\lim _{n \rightarrow \infty}
\ket{\Phi _n} = \ket{\Phi }$. Moreover, if we add one condition;
$\limsup _{n \rightarrow \infty} p_n >0$, then we can show the
contradiction as follows.

Since $\limsup _{n \rightarrow \infty} p_n >0$, there exists a
subsequence of $p_{n}$, such that $\lim p_{n(k)} = p >0$ and
$p_{n(k)} >p/2$ for all $k \in \mathbb{N}$. Then, since $\ket{\Psi
}$ is $\epsilon$-convertible to $\ket{\Phi _n}$ with probability
$p_n$, for any $\epsilon >0$ and for any $k \in \mathbb{N}$, there
exists an SLOCC operation $\Lambda _{\epsilon,  n(k)}$ such that
$\| \Lambda_{\epsilon,  n(k)}(\ket{\Psi}\bra{\Psi})/\Tr
\Lambda_{\epsilon,  n(k)}(\ket{\Psi}\bra{\Psi}) - \ket{\Phi
_{n(k)}}\bra{\Phi _{n(k)}} \| < \epsilon$ and $\Tr
\Lambda_{\epsilon,  n(k)} (\ket{\Psi}\bra{\Psi}) \ge p_{n(k)} >
p/2$. Moreover, since $\lim _{n \rightarrow \infty}\ket{\Phi _n}
=\ket{\Phi}$, for any $\epsilon >0$, there exists an $N_{\epsilon}
\in \mathbb{N}$ such that for any $n \ge N_{\epsilon}$, $\|
\ket{\Phi _n}\bra{\Phi _n} - \ket{\Phi}\bra{\Phi} \| < \epsilon$.
Therefore, for any $2 \epsilon >0$, by choosing $k \in \mathbb{N}$
as $n(k) \ge N_{\epsilon}$,
\begin{eqnarray*}
&\quad & \| \Lambda _{\epsilon, n(k)}/\Tr \Lambda_{\epsilon, n(k)}
-
\ket{\Phi}\bra{\Phi} \| \\
&\le & \| \Lambda_{\epsilon, n(k)}/\Tr \Lambda_{\epsilon, n(k)} -
\ket{\Phi _{n(k)}}\bra{\Phi _{n(k)}} \| + \| \ket{\Phi
_{n(k)}}\bra{\Phi _{n(k)}} - \ket{\Phi}\bra{\Phi}\| \\
&\le & 2 \epsilon.
\end{eqnarray*}
Moreover, $\Tr \Lambda_{\epsilon, n(k)}(\ket{\Psi}\bra{\Psi}) \ge
p_{n(k)} >p/2$. Therefore, $\ket{\Psi}$ is $\epsilon$-convertible
to $\ket{\Phi}$ with probability $p/2$. This is a contradiction.
Therefore, if $\ket{\Psi}$ is not $\epsilon$-convertible to
$\ket{\Phi} $, and if $\ket{\Psi}$ is $\epsilon$-convertible to
$\ket{\Phi _n}$ with probability $p_n$ where $\lim _{n \rightarrow
\infty} \ket{\Phi_n} = \ket{\Phi}  $, then, $\lim _{n \rightarrow
\infty} p_n = 0$. \hfill $square$
\end{Proof}
This lemma means that if $\ket{\Psi}$ cannot be converted to
$\ket{\Phi}$ by SLOCC, then $\ket{\Psi}$ is also almost certainly
inconvertible to states near to $\ket{\Phi}$. Therefore, our
definition of $\epsilon$-convertibility preserves continuity of
the theory (at least sufficiently for the purposes of the
classification of states), and we can avoid the discontinuity
difficulty mentioned above.


\subsection{Nielsen's and Vidal's theorems for infinite-dimensional systems}
\label{nielsen vidal}

In this subsection we reconstruct Nielsen's and Vidal's theorems
for infinite-dimensional systems by means of
$\epsilon$-convertibility. As a result, we will see that we can
also avoid the difficulty of a potentially infinite cost for
classical communication by our convertibility, that is, only a
finite amount of classical communication is actually necessary for
our theory of convertibility. As is well known, Nielsen's and
Vidal's theorems give the necessary and sufficient conditions of
LOCC and SLOCC, respectively. Therefore, by proving these theorems
rigorously we may obtain a firm foundation for the analysis of
SLOCC convertibility for infinite-dimensional systems, which we
shall consider in the next section. Since Vidal's theorem is a
generalization of Nielsen's theorem, we shall first discuss
Nielsen's theorem and then go on to consider Vidal's theorem.

In finite-dimensional systems Nielsen's theorem gives the
necessary and sufficient conditions for LOCC convertibility
between a pair of bipartite pure states $\ket{\Phi}$ and
$\ket{\Psi}$ as follows
\begin{equation}\label{nielsenfinite}
     \ket{\Psi} \rightarrow \ket{\Phi}
     ~~ \Leftrightarrow ~~
     \mathbf{\lambda} \prec \mathbf{\mu}\;,
\end{equation}
where the arrow $\rightarrow$ represents convertibility under
LOCC, and $\mathbf{\lambda}$ and $\mathbf{\mu}$ represent
sequences of Schmidt coefficients (in descending order) of the
states $\ket{\Psi}$ and $\ket{\Phi}$, respectively, and $\prec$
denotes majorization of the sequences \cite{majorization} (if
$\mathbf{\lambda} \prec \mathbf{\mu}$, we say ``$\mathbf{\lambda}$
is majorized by $\mathbf{\mu}$'').  In infinite-dimensional
systems we can show that Eq.~(\ref{nielsenfinite}) is still valid
where we replace the meaning of $\rightarrow$ by
$\epsilon$-convertibility under LOCC. Nielsen's theorem then takes
the following form for infinite-dimensional systems:
\begin{Theorem} \label{epsilon Nielsen}
$\ket{\Psi}$ is $\epsilon$-convertible to $\ket{\Phi}$, if and
only if $\lambda \prec \mu $, where $\prec$ means majorization in
infinite-dimensional systems (see Appendix \ref{Majorization}),
and $\mathbf{\lambda}$ and $\mathbf{\mu}$ are the Schmidt
coefficients of $\ket{\Psi}$ and $\ket{\Phi}$, respectively.
\end{Theorem}
Since the proof of theorem \ref{epsilon Nielsen} is long, we have
placed the rigorous proof of this theorem in Appendix \ref{Proof
of Nielsen}. Below we only give a sketch of the proof:

\textbf{Sketch of Proof}

1) The necessary part: We can directly extend the proof of
necessity of the original theorem to infinite-dimensional systems.
The necessary condition part of the original theorem is
constructed using the Lo-Popescu theorem (Theorem
\ref{Lo-Popescu}) \cite{lo-popescu} and Uhlmann's theorem (Theorem
\ref{Uhlmann}). Since these two theorems can themselves be
extended to infinite-dimensional systems (see Appendix A and B).
The same proof for finite-dimensional systems still holds in
infinite-dimensional systems.

2) The sufficient part: In the proof of sufficiency, our
definition of $\epsilon$-convertibility plays a crucial role in
extending the proof of Nielsen's theorem. Our proof is based on
the proof for finite-dimensional systems in Ref.~\cite{uniqueness}
and is extended to genuine infinite-dimensional states by means of
$\epsilon$-convertibility. We can show that for any $N$, there
exists a state $ \ket{\Phi'}$ (which depends on $N$)
such that its first $N$ Schmidt coefficients are equal to the
Schmidt coefficients of $\ket{\Phi}$ and where the Schmidt
coefficients of $\ket{\Psi }$ are majorized by the Schmidt
coefficients of $\ket{\Phi'}$. Therefore, for every neighborhood of
$\ket{\Phi}$, we can always find a state to which $\ket{\Psi}$ can
be converted under LOCC. \hfill $\square$

By means of Nielsen's theorem in infinite-dimensional systems we
can extend Vidal's theorem for SLOCC convertibility \cite{vidal},
which gives the necessary and sufficient condition of SLOCC
convertibility with a probability $p$ to infinite-dimensional
systems using $\epsilon$-convertibility. Vidal's theorem states
that a bipartite pure state $\ket{\Psi}$ can be converted to
another bipartite pure state $\ket{\Phi}$ under SLOCC with
probability at least $p$ if and only if $\lambda\prec ^{\omega} p
\mu$ [here $\prec ^{\omega}$ denotes super-majorization and is defined
in Appendix A1, Eq. (\ref{super-majorization}) of Definition 4].
The generalization of Vidal's theorem for
$\epsilon$-convertibility can then be written:
\begin{Theorem} \label{epsilon Vidal}
$\ket{\Psi}$ is $\epsilon$-convertible to $\ket{\Phi}$ by SLOCC
with probability $p$, if and only if $\lambda \prec ^{\omega} p
\mu$ are satisfied where $\lambda$ and $\mu$ are the Schmidt
coefficients of $\ket{\Psi}$ and $\ket{\Phi}$ respectively.
\end{Theorem}
\begin{Proof}
The proof of this theorem is in appendix \ref{Proof of Vidal}.
\end{Proof}
Therefore, the extension of Vidal's theorem also applies to
$\epsilon$-convertibility.

Although infinite amounts of classical information do not exist in
the real world,  an infinite amount of classical communication is
necessary to convert one genuine infinite-dimensional state to
another by LOCC and SLOCC in the conventional theory of
convertibility. From the proof of Theorem \ref{epsilon Nielsen},
we can show that we can avoid such infinite costs of classical
communication in LOCC convertibility by our new definition of
$\epsilon$-convertibility. In the proof of this theorem, we showed
that there exists a natural number $M$ such that $\ket{\Phi'}$
satisfies the condition $\mu'_N = \lambda _N$ for $N \ge M$.
Therefore, the LOCC operation by which $\ket{\Psi }$ can be
converted into $\ket{\Phi'}$ is actually an LOCC operation
requiring only a finite amount of classical communication.  We can
also show a similar result for SLOCC convertibility. By the proof
of Theorem \ref{epsilon Vidal} it is easily seen that we can
construct the protocol of SLOCC with only a finite amount of
classical communication in a manner similar to LOCC. As a result,
in our definition of $\epsilon$-convertibility of LOCC and SLOCC,
we can convert states with any finite accuracy by only a finite
amount of communication, and only when this error goes to zero
does the amount of classical communication go to infinity.
Therefore, our definition of $\epsilon$-convertibility yields a
theory of single-copy LOCC and SLOCC convertibility requiring only
a finite amount of classical communication even in the
infinite-dimensional setting.

Here we need to add two final remarks about our framework of
$\epsilon$-convertibility. From the proofs of Theorems
\ref{epsilon Nielsen} and~\ref{epsilon Vidal}, we can derive
another interpretation of $\epsilon$-convertibility. First, in the
case of LOCC, that is, Nielsen's theorem, since the state
$\ket{\Phi '} = \sum _{k=1}^{\infty} \sqrt{\mu '} \ket{i} \otimes
\ket{i}$ is also majorized by $\ket{\Phi}$ in the proof of Theorem
\ref{epsilon Nielsen} (Appendix \ref{Proof of Nielsen}), we can
immediately see the following fact: If $\lambda \prec \mu $ where
$\lambda$ and $\mu$ are the Schmidt coefficients of $\ket{\Phi}$
and $\ket{\Psi}$ respectively, then there exists a sequence of
LOCC $\{ \Lambda _n \} _{n=1}^{\infty}$ such that for all $n \in
\mathbb{N}$, $\Lambda_n \cdots \Lambda _1 (\ket{\Psi} \bra{\Psi}
)$ has the same Schmidt basis and their Schmidt coefficients are
majorized by those of $\ket{\Phi}$ and they also satisfy $\lim _{n
\rightarrow \infty} \Lambda _n \cdots \Lambda _1 (\ket{\Psi}
\bra{\Psi} ) = \ket{\Phi} \bra{\Phi}$. Thus, we can interpret the
above sequence of LOCC as the LOCC with an infinite number of
steps of classical communication. Second, since in the proof of
the above theorem in appendix \ref{Proof of Vidal}, we constructed
a sequence of SLOCC $\{ \Lambda _n \}_{n=1}^{\infty}$ such that
$\lim _{n \rightarrow \infty} \Lambda _n \cdots \Lambda _1
(\ket{\Psi} \bra{\Psi} )/\Tr\; \Lambda _n \cdots \Lambda _1
(\ket{\Psi} \bra{\Psi} ) = \ket{\Phi}\bra{\Phi}$, we can consider
that Vidal's theorem is also naturally extended to
infinite-dimensional systems by the redefinition of LOCC
convertibility including an infinite number of steps of classical
communication. Therefore, we can also say that both the Nielsen
and Vidal theorems can be extended to infinite-dimensional systems
if we allow for an infinite number of steps of LOCC.

In this section, we proposed a new definition of convertibility,
{\itshape $\epsilon$-convertibility}, to treat entanglement
convertibility between genuine infinite-dimensional states. This
redefinition is suitable from both the technical and realistic
viewpoints, that is, to avoid the difficulties of both
discontinuity and infinite cost in classical communication for
infinite-dimensional systems. Then, by means of
$\epsilon$-convertibility we proved the Nielsen and Vidal theorems
which are the fundamental theorems of LOCC and SLOCC
convertibility in infinite-dimensional systems. As a result under
our change of definition the framework of entanglement
convertibility is preserved in the context of infinite-dimensional
systems, and therefore our definition of $\epsilon$-convertibility
for LOCC is suitable and sufficient for realistic conditions of
quantum information processing in infinite-dimensional systems.


\section{Extension of Schmidt rank} \label{Extension}
\subsection{Definition and its basic property} \label{definition}

In this section we discuss the SLOCC convertibility of
infinite-dimensional systems and show that there are many
important differences between the structure of the SLOCC
classification of genuinely infinite-dimensional states and that
of finite-dimensional states. For this purpose in this subsection
we first define a pair of new SLOCC monotones which can be
considered as extensions of the Schmidt rank, and then we analyze
their properties. Finally, we show that there are continuously
many classes of states under SLOCC convertibility in
infinite-dimensional systems. In the following we always consider
SLOCC convertibility in the meaning of $\epsilon$-convertibility
defined above. Therefore, we henceforth omit the qualifier
`$\epsilon$'.

To study convertibility in detail, monotones of convertibility are
crucially important. In finite systems, the Schmidt rank (the rank
of the reduced density matrix) gives the necessary and sufficient
condition of SLOCC convertibility. On the other hand, in
infinite-dimensional systems, since almost all states have
infinite Schmidt rank, the classification by Schmidt rank is not
useful. Therefore, proposals for new SLOCC monotones for genuine
infinite-dimensional entangled states are essential for analysis
of SLOCC convertibility between genuinely infinite-dimensional
states. In the followings, we give a definition of a pair of new
SLOCC monotones $R^-$ and $R^+$, which can be considered an
extension of the Schmidt rank for infinite-dimensional systems.

Since the usual Schmidt rank represents how quickly Schmidt
coefficients vanish, when we consider their extension to genuine
infinite-dimensional states it is natural to define the extension
of this concept as a function which represents how quickly a
sequence of Schmidt coefficients converge to zero. In Vidal's
theorem Schmidt coefficients always appear in the form of a sum
from $n$ to $\infty$ which is an LOCC monotone for all $n \in
\mathbb{N}$ and is called ``{\it Vidal's monotone}''
\cite{monotone}. Therefore, rather than studying the direct
convergence of Schmidt coefficients $\{ \lambda _n
\}_{n=1}^{\infty}$ we shall study the convergence of Vidal's
monotones $\{ \sum _{i=n}^{\infty} \lambda _i \}_{n=1}^{\infty}$.
To measure the speed of convergence of Vidal's monotone we compare
a sequence of Vidal's monotones with some real parameterized class
of sequences. Thus, we define the new monotones as follows:
\begin{Definition} \label{def of monotone}
For a parameterized class of sequences $\{ f_r (n) \} _{r \in
(a,b)}$ which satisfy the following conditions:
\begin{list}{}{}
     \item $\forall r \in (a,b)$, $\lim _{n \rightarrow \infty} f_r(n) =0$
         \item $\forall r \in (a,b)$,
               $n_1 < n_2 \Rightarrow f_r(n_1) > f_r(n_2) > 0$
         \item $\forall n \in \mathbb{N}$,
               $r_1 < r_2 \Rightarrow \lim _{n \rightarrow \infty}
                    \frac{f_{r_1} (n) }{f_{r_2} (n) } =0$\;,
\end{list}
where $n \in \mathbb{N}$, $0 \le a < b \le \infty$, we define a
pair of functions $R^+_{f_r} (\ket{\Psi} )$ and $R^-_{f_r}
(\ket{\Psi})$ by
\begin{eqnarray}
R^+ _{f_r}(\ket{\Psi}) &=& \inf \{ r \in (a,b) | \lim _{n
\rightarrow \infty}
  \frac{\sum _{i=n}^{\infty} \lambda _i}{f_r(n)} = 0 \} \\
R^- _{f_r}(\ket{\Psi}) &=& \inf \{ r \in (a,b) | \underline{\lim}
_{n \rightarrow \infty}
  \frac{\sum _{i=n}^{\infty} \lambda _i}{f_r(n)} = 0 \}\;.
\end{eqnarray}
If for all $r \in (a,b) $, $\overline{\lim} _{n \rightarrow
\infty} {\sum _{i=n}^{\infty} \lambda _i}/{f_r(n)} > 0$, then we
define $R^+_{f_r} (\ket{\Psi}) =b$. Here, we use the notation of
$\overline{\lim} = \limsup$ and $\underline{\lim} = \liminf$.
\end{Definition}
For the definition of $R^+_{f_r}(\ket{\Psi})$, we could also have
defined this function as
\begin{equation*}
R^+ _{f_r}(\ket{\Psi}) = \inf \{ r \in (a,b) | \overline{\lim} _{n
\rightarrow \infty}
  \frac{\sum _{i=n}^{\infty} \lambda _i}{f_r(n)} = 0 \}\;,
\end{equation*}
however this definition is the same as the previous definition, since
$\overline{\lim} _{n \rightarrow \infty} {\sum _{i=n}^{\infty}
\lambda _i}/{f_r(n)} = 0$ guarantees
  $\lim_{n \rightarrow \infty}
{\sum _{i=n}^{\infty} \lambda _i}/{f_r(n)} = 0$.
Note that the limits
$\overline{\lim} _{n \rightarrow \infty} \sum _{i=n}^{\infty}
\lambda _i / f_r(n)$ and $\underline{\lim} _{n \rightarrow \infty}
\sum _{i=n}^{\infty} \lambda _i /f_r(n)$ do not generally
coincide. Thus, to measure the speed of the convergence, we need
two functions $R^+_{f_r}(\ket{\Psi})$ and $R^-_{f_r}(\ket{\Psi})$
corresponding to these different approaches of the limit as given
above. By their definition, we can easily see that $R^+ _{f_r}$
and $R^- _{f_r}$ satisfy $R^- _{f_r} (\ket{\Psi}) \le R^+ _{f_r}
(\ket{\Psi})$ for all $\ket{\Psi}$. As we might expect, both $R^+
_{f_r}$ and $R^- _{f_r}$ are SLOCC monotones, and moreover, the
sufficient condition of monotonicity is also partially valid as
given by the statement of the following theorem. (Note, that when
the choice of $f_r(x)$ is clear we write simply $R^+$ and $R^-$
for the monotones.)
\begin{Theorem} \label{theorem of monotone}
For all $f_r$ which satisfy the  condition in Definition \ref{def
of monotone},
\begin{enumerate}
\item If $\ket{\Psi}$ can be converted to $\ket{\Phi}$ by SLOCC then
$R^+_{f_r} (\ket{\Psi}) \ge R^+_{f_r} (\ket{\Phi})$ and $R^-_{f_r}
(\ket{\Psi}) \ge R^-_{f_r} (\ket{\Phi})$. \item If
$R^+_{f_r}(\ket{\Phi}) < R^-_{f_r}(\ket{\Psi})$, then $\ket{\Psi}$
can be converted to $\ket{\Phi}$ by SLOCC.
\end{enumerate}\end{Theorem}
\begin{Proof}
Proof of part 1: \\
We only prove this for the case of $R^+ _{f_r}$ since the proof
for $R^- _{f_r}$ is identical. Suppose $R^+ _{f_r} (\ket{\Phi}) >
R^+ _{f_r} (\ket{\Psi})$ then for all $R^+ _{f_r} (\ket{\Psi}) < r
< R^+ _{f_r} (\ket{\Phi})$, $\overline{\lim} _{n \rightarrow
\infty}  \sum _{i=n}^{\infty} \lambda _i/f_r (n) = 0$ and \newline
$\overline{\lim} _{n \rightarrow \infty}  \sum _{i=n}^{\infty} \mu
_i / f_r (n) > 0$, where $\{ \lambda _i \}_{i=0}^{\infty}$ and $\{
\mu _i \}_{i=0}^{\infty}$ are Schmidt coefficients of $\ket{\Psi
}$ and $\ket{\Phi }$, respectively. Thus, for all $\delta
> 0$ there exists an $N_0 (\delta)$ such that if $n > N _0 (\delta
)$, then $\sum _{i=n}^{\infty} \lambda _i / f_r(n) < \delta $.
Suppose $a \stackrel{def}{=} \overline{\lim} _{n \rightarrow
\infty} \sum _{i=n}^{\infty} \mu _i / f_r(n) >0$, then there
exists a partial sequence of $\sum _{i=n}^{\infty} \mu _i /
f_r(n)$, say $\sum _{i=k(n)}^{\infty} \mu _i / f_r(k(n))$, such
that $\lim _{n \rightarrow \infty} \sum _{i=k(n)}^{\infty} \mu _i
/ f_r(k(n)) = a
>0$. Then there exists an $N_1 \in \mathbb{N}$ such that for all $n > N_1$,
$\sum _{i=k(n)}^{\infty} \mu _i / f_r(k(n)) > a/2$. Therefore if
we define $N_2(\delta)$ as $N_2(\delta) = \max (N_1, \min \{n \in
N |k(n) \ge N_0 \})$, then for all $n > N_2(\delta)$, $\sum
_{i=k(n)}^{\infty} \lambda _i / f_r(k(n)) < \delta$ and $
f_r(k(n)) / \sum _{i=k(n)}^{\infty} \mu _i < 2/a$. That is, $\sum
_{i=k(n)}^{\infty} \lambda _i / \sum _{i=k(n)}^{\infty} \mu _i <
2\delta / a$. This means $\underline{\lim}_{n \rightarrow
\infty} \sum _{i=n}^{\infty} \lambda _i / \sum _{i=n}^{\infty} \mu
_i =0$,  which means $\ket{\Psi }$ cannot be convertible to
$\ket{\Phi }$ by SLOCC from Vidal's Theorem.

Proof of part 2: \\
If $R^+(\ket{\Phi}) < R^-(\ket{\Psi})$, then for all
$R^+(\ket{\Phi}) < r < R^-(\ket{\Psi})$, $a \stackrel{def}{=}
\underline{\lim} _{n \rightarrow \infty} \sum _{i=n}^{\infty}
\lambda _i / f_r(n) >0$ and $\lim _{n \rightarrow \infty} \sum
_{i=n}^{\infty} \mu _i / f_r(n) =0$. Then, for all $\delta > 0$
there exists an $N_0(\delta)$  such that if $n > N_0(\delta)$,
$\sum _{i=n}^{\infty} \lambda _i / \sum _{i=n}^{\infty} \mu _i
> a / 2\delta$. That is $\lim _{n \rightarrow \infty} \sum
_{i=n}^{\infty} \lambda _i / \sum _{i=n}^{\infty} \mu _i =
\infty$. From Vidal's theorem, this means that $\ket{\Psi }$ can
be converted to $\ket{\Phi }$ by SLOCC. \hfill $\square$
\end{Proof}
Hence, 
this SLOCC monotone satisfies the sufficient condition of
convertibility of SLOCC at least with the above meaning. As we
shall see in the following sections by using
$R^+_{f_r}(\ket{\Psi})$ and $R^-_{f_r}(\ket{\Psi})$ together, we
can determine the classification of SLOCC convertibility better
than in the case of using the other SLOCC monotones, although we
also need both $R^+_{f_r}$ and $R^-_{f_r}$ to lead to the
sufficient condition. Therefore, we can consider this pair of
monotones as extensions of the Schmidt rank.

In the last part of this subsection we note one important fact
which we can easily see from Theorem \ref{theorem of monotone},
that is that ``{\it in infinite dimensional systems there are at
least continuously infinitely many different classes of SLOCC
convertibility.}'' Since $R^+(\ket{\Psi})$ (or $R^- (\ket{\Psi})$)
is an SLOCC monotone whose range is a non-trivially connected set
(interval) of real numbers, if $\ket{\Psi _r}$ satisfies
$R^+(\ket{\Psi _r}) =r$, each $\ket{\Psi _r}$ should belong to
different classes of SLOCC convertibility for every different
value of $r$. That is, there exists an injective map from a
non-trivially connected set of real numbers to the quotient set of
states by SLOCC. Therefore, in infinite-dimensional bipartite
systems, the cardinal number of the quotient set of states by
SLOCC convertibility is greater than or equal to the cardinal
number of the {\it continuum}, (where the cardinal number of the
continuum is equal to the cardinal number of an arbitrary interval
of real numbers)\cite{kolmogorov}. Comparing this to the
finite-dimensional systems case, where the cardinal number of the
quotient set of states by SLOCC convertibility is equal to the
dimension of the local systems. This fact is remarkable, that is,
the cardinal number of such classes is actually larger than the
local dimension (which is only {\it countably infinite}) in
infinite-dimensional systems.

\subsection{Examples of $R^+(\ket{\Psi})$ and $R^-(\ket{\Psi})$}
\label{exsample}

In this subsection we construct important examples of the SLOCC
monotones $R^+(\ket{\Psi})$ and $R^-(\ket{\Psi})$, and analyze
SLOCC convertibility between some interesting classes of genuinely
infinite-dimensional states. One is a class of states with
polynomially-damped Schmidt coefficients; another is the class of
two-mode squeezed states. Since these new monotones depend on a
real parameterized family of sequences $\{ f_r(n) \}_{r \in
(a,b)}$, we need to choose this family suitably to analyze SLOCC
convertibility among particular states. For this purpose it is
convenient to derive the reference-states class $\{ \ket{\Psi _r}
\}_{r \in (a,b)}$ for particular $\{ f_r(n) \}_{r \in (a,b)}$, as
the states which satisfy the condition $R^+(\ket{\Psi _r}) =
R^-(\ket{\Psi _r}) =r$. Therefore, at first we construct a way of
finding the reference class $\ket{\Psi _r}$ from $f_r(n)$. The
following corollary gives a method.
\begin{Corollary} \label{reference}
If $\{ f_r (n) \} _{r \in (a,b), n \in \mathbb{N}}$ satisfies
following conditions:
\begin{enumerate}
     \item $\forall r \in (a,b), n_1 \le n_2 \Rightarrow
  f_r(n_1) > f_r(n_2)$ (monotonically decreasing)
               \label{monotonically decreasing}
         \item $\forall r \in (a,b)$ and  $\forall n \in \mathbb{N},
               f_r(n) + f_r(n+2) \ge 2f_r(n+1)$ (convexity) \label{convexity}
         \item $\forall m \in \mathbb{N}$, $r_1 \le r_2 \Leftrightarrow
               \lim _{n \rightarrow \infty}
               \frac{f_{r_1}(n)}{f_{r_2}(n+m)} =0$ (monotonicity)\;,
               \label{monotonisity}
\end{enumerate}
then, $\ket{\Psi _r} = \frac{1}{c_r} \sum _{n=1}^{\infty} \sqrt{-
f^{'}_r (n)} \ket{n} \otimes \ket{n}$, where $c_r= \sum _{n
=1}^{\infty} -f^{'}_r(n)$, satisfies $R^+(\ket{\Psi _r}) =
R^-(\ket{\Psi _r}) = r$ which are made from $\{ f_r (n) \} _{r \in
\mathbb{R}, n \in \mathbb{N}}$, and where $f_r'(x)$ denotes the
derivative of $f_r(x)$ with respect to $x$.

\end{Corollary}
\begin{Proof}
 From conditions \ref{monotonically decreasing} and \ref{convexity}
above, there exists a class of doubly differentiable functions $\{
f_r(x) \} _{r \in (a,b), x \in \mathbb{R^+ }}$ which are an
extension of the sequences $\{ f_r (n) \} _{r \in (a,b), n \in
\mathbb{N}}$ such that they satisfy $f^{'}_r(x) < 0$ and
$f^{''}_r(x) \ge 0$. Therefore, a class of states $\{ \ket{\Psi
_r} \}_{r \in (a,b)}$ is well defined and their Schmidt
coefficients are $\{ -f^{'}_r(n) /c_r \} _{n=1}^{\infty}$ in
decreasing order. By definition then
\begin{eqnarray}
\int _{n}^{\infty} \frac{-f^{'}_r(x)}{c_r} dx & \le & \sum
_{k=n}^{\infty} \frac{-f^{'}_r(n)}{c_r}  \le  \int _{n-1}^{\infty}
\frac{-f^{'}_r(x)}{c_r} dx
\nonumber \\
\frac{f_r(n)}{c_rf_{r_1}(n)} & \le & \sum _{k=n}^{\infty}
\frac{-f^{'}_r(n)}{c_rf_{r_1}(n)}  \le
\frac{f_r(n-1)}{c_rf_{r_1}(n)}\;. \nonumber
\end{eqnarray}
If $r<r_1$ then
\begin{equation}
\lim _{n \rightarrow \infty} \sum _{k=n}^{\infty}
\frac{-f^{'}_r(n)}{c_rf_{r_1}(n)} \le \lim _{n \rightarrow \infty}
\frac{f_r(n-1)}{c_rf_{r_1}(n)} = 0\;,\nonumber
\end{equation}
and if $r>r_2$ then
\begin{equation}
\lim _{n \rightarrow \infty} \sum _{k=n}^{\infty}
\frac{-f^{'}_r(n)}{c_rf_{r_1}(n)} \ge \lim _{n \rightarrow \infty}
\frac{f_r(n)}{c_rf_{r_1}(n)} = +\infty \;.\nonumber
\end{equation}
Thus, $R^+(\ket{\Psi _r}) = R^-(\ket{\Psi _r}) = r$. \hfill
$\square$
\end{Proof}
This Corollary means that with the above three additional
conditions for $f_r(n)$ we may always derive a class of reference
states which correspond to each value of $R^-(\ket{\Psi})$ and
$R^+(\ket{\Psi})$.

In what follows we construct examples of SLOCC monotones by means
of the above Corollary and analyze two remarkable classes of
states. One corresponds to the states which belong a higher class
of SLOCC convertibility and the other to the well-known two-mode
squeezed states.

As a first example consider $R^-(\ket{\Psi})$ and
$R^+(\ket{\Psi})$ made from $\{ f_r(n) = n^{-(\frac{1}{r}-1)} \}
_{r \in (0,1)}$. By Corollary \ref{reference}, their class is
\begin{equation}
\ket{\Psi _r} = \frac{1}{\sqrt{\zeta (1/r)}} \sum _{n=1}^{\infty}
  n^{-\frac{1}{2r}}  \ket{n} \otimes \ket{n}\;,
\end{equation}
where $\zeta (x)$ is the Riemann zeta function as a normalization
factor. By definition, $R^-(\ket{\Psi})$ and $R^+(\ket{\Psi})$
represent how quickly the Schmidt coefficients of $\ket{\Psi}$
converge to $0$ as a polynomially-damped function. Thus, this
function has a strictly positive value for states with
polynomially-damped Schmidt coefficients. Similarly, for all
states $\ket{\Psi}$ for which the Schmidt coefficients damp
exponentially like two-mode squeezed states, we have
$R^-(\ket{\Psi}) = R^+(\ket{\Psi}) =0$. Because the Schmidt
coefficients can never be proportional to $1/n$ asymptotically in
infinite-dimensional systems (since $\sum _{n=1}^{\infty} 1/n =
\infty$), for small $\epsilon > 0$, $\ket{\Psi _r}$ with $r=1-
\epsilon$ can be converted to almost any state. In the above sense
we can say that they belong to a ``higher rank'' of entanglement
class in terms of single-copy SLOCC. On the other hand, the above
state with small $\epsilon$ is not of the ``{\it highest
states}''. That is, we can consider a class of states which belong
to a higher rank of entanglement class than $\{ \ket{\Psi _r}
\}_{r \in (0,1)}$ as follows.  For the states
\begin{equation}
\ket{\Psi _t} = \frac{1}{C_t} \sum_{n=1}^{\infty}
\frac{1}{\sqrt{x(\log x)^t}} \ket{n} \otimes \ket{n}\;,
\end{equation}
with $t>0$, $R^-(\ket{\Psi _t}) = R^+(\ket{\Psi _t})=1$, and we
can easily see that for all $t >0$, $\ket{\Psi _r}$ can not be
converted to $\ket{\Psi _t}$ by SLOCC. In a similar manner, for all
one-parameter classes of states we can always define a class of
states which belong to a higher rank and can define a new pair of
monotones from this class of states. Therefore, there does not
exist a highest one-parameter class of states within the SLOCC
classification.

As a next example consider $f_q(n) = e^{2n \log q} = q^{2n}$, $q
\in (0,1)$. In this case, the reference class is $\ket{\Psi _q}=
\frac{1}{c_q} \sum _{n=1}^{\infty} q^n \ket{n} \otimes \ket{n}$,
that is the well-known two-mode squeezed states with
$\frac{1}{2}\log \frac{1+q}{1-q}$ being the squeezing parameter.
Therefore,  $R^+_{f_q}(\ket{\Psi})$ and
$R^-_{f_q}(\ket{\Psi})$ can be regarded as being analogs of
squeezing parameters for any entangled states.

The above two examples also show that the classification of SLOCC
is quite different from the classification by the amount of
entanglement, that is, the classification of asymptotic
(infinite-copy) LOCC in infinite-dimensional systems. In
infinite-dimensional systems we often consider the class of
two-mode squeezed states
$\ket{\Psi_q}=\frac{1}{c_q}\sum_{n=1}^{\infty}q^n\ket{n}\otimes\ket{n}$
with $q = 1 - \epsilon$ instead of the maximally entangled states.
Because $ \lim _{q \rightarrow 1} E(\ket{\Psi _q}) = \infty$ this
state converts to almost any state asymptotically by infinite-copy
LOCC with unit probability. However by single-copy SLOCC with
non-zero probability they cannot be converted to states
$\ket{\Psi}$ with $R^+(\ket{\Psi})>0$, where the monotone
$R^+(\ket{\Psi})$ is made from $f_r (n) = n^{-(\frac{1}{r}-1)}$.
On the other hand, if we consider the class of states $\ket{\Psi
_r}= \sum _{n=1}^{\infty} n^{-\frac{1}{2r}}  \ket{n} \otimes
\ket{n}/\sqrt{\zeta (1/r)}$ as we have already seen for small
$\epsilon >0$, $\ket{\Psi _{1-\epsilon}}$ can be converted to
almost any state by single-copy SLOCC with non-zero probability.
Although the amount of entanglement for both $\{ \ket{\Psi _q} \}$
and $\{ \ket{\Psi _r } \}$ tend to infinity in the limit, $\{
\ket{\Psi _r} \}$ belongs to a higher class than $\{ \ket{\Psi _q}
\}$ in the single-copy scenario.

We add one final remark here: Although, we have only presented two
examples for $f_r(n)$, there may be many other examples which are
important in some situations. Generally speaking, for any given
states, we can find a suitable function $f_r(n)$ for the analysis
of the states. For example, if we deal with states whose Schmidt
coefficients damp exponentially we can chose $f_r(n) = \exp
(n^{-1/r}), \exp (\exp (n^{-1/r})), $ etc, as the
coefficients damp quickly enough to evaluate the monotones for the
states.

\subsection{Strong inhibition law}
\label{inhibition law}

So far we have emphasized the difference between states with
exponentially-damped Schmidt coefficients and those with
polynomially-damped coefficients and shown that
exponentially-damped states cannot be converted into
polynomially-damped states no matter how large their measure of
entanglement is. Here we shall give one more fact which will
demonstrate the remarkable difference between exponentially and
polynomially-damped states. That is, ``\textit{However finitely
many copies there are, exponentially-damped states cannot be
converted into polynomially-damped states}.'' This fact can be
showed as the follows: Suppose $\ket{\Psi}$ is an
exponentially-damped state and $\ket{\Phi}$ is a
polynomially-damped one, then rigorously speaking, there exists a
real number $r$ and a polynomial $p(n)$ which satisfy $\lim _{n
\rightarrow \infty} \frac{g_{\ket{\Psi}}(n)}{e^{-rn}}=0$ and
$\underline{\lim}_{n\rightarrow \infty}
\frac{p(n)}{g_{\ket{\Phi}}(n)} =0$, where $g_{\ket{\Psi}}(n)$ is
Vidal's monotone of $\ket{\Psi}$. Define $\ket{\xi _r} =
\frac{1}{C_r} \sum _{n=1}^{\infty} e^{-rn} \ket{n}\otimes\ket{n}$,
then we have
\begin{eqnarray}
\ket{\xi _r} ^{\otimes p}&=& \frac{1}{C_r^p} \sum _{n_1, n_2,
\cdots ,n_p}^{\infty} e^{-r(n_1+n_2+ \cdots +n_p)} \ket{n_1, n_2,
\cdots , n_p}\otimes \left \vert n_1, n_2,\cdots,n_p\right\rangle
\nonumber \;.\!\!
\end{eqnarray}
If we reorder the Schmidt terms to the form $\ket{\xi _r}
^{\otimes p} = \frac{1}{C} \sum _{k=1}^{\infty} f(k)
\ket{k}\otimes \ket{k}$ we can see by easy calculation that $f(k)
\le e^{-r[ (p! k )^{1/p} +1]}$. Thus, we have $\lim _{n
\rightarrow \infty} {f(n)}/{p(n)} =0$ and this means $\ket{\Psi}
^{\otimes p}$ cannot converted into $\ket{\Phi}$ for any $p \in
\mathbb{N}$. This result shows that in infinite-dimensional
systems some classes of states (like states with finite Schmidt
ranks, with exponentially-damped Schmidt coefficients, and with
polynomially-damped Schmidt coefficients) can be distinguished
from each other more strongly than the case of finite-dimensional
systems by SLOCC classification. Thus, \textit{with arbitrary
finitely-many copies}, we also cannot convert the states from
finite rank to infinite rank, and similarly from exponentially
damped to polynomially damped. In finite-dimensional systems,
there is no feature like this. Therefore, these properties of
entanglement are genuine for infinite-dimensional systems and show
the special strong position of states with polynomially-damped
Schmidt coefficients from the view of finite-copy transformations.

As a final remark for this section we must discuss the energy of
such long-tailed states. In realistic situations the set of states
which can be produced experimentally will be limited by some bound
in energy. Therefore, it is essential to consider the subset of
states which consist of states restricted to that bounded energy.
However, for several states with polynomially-damped Schmidt
coefficients, the mean value of a polynomial Hamiltonian, like for
example the harmonic oscillator, diverges.  Therefore, generally
only a fraction of polynomially-damped states can be created in
laboratories.


\section{Summary}
\noindent
In this paper in order to avoid the difficulties of discontinuity
and infinite amounts of classical communication in the theory of
SLOCC convertibility of infinite-dimensional systems, we proposed
a new definition of convertibility, {\it
$\epsilon$-convertibility}, as the convertibility of states in an
approximated setting by means of the trace norm. In the Section
\ref{infinite Nielsen} we showed that this definition guarantees
at least weak continuity for SLOCC convertibility (Lemma
\ref{epsilon SLOCC lemma}), and guarantees that the protocol only
uses finite amounts of classical communication. Then, we
reconstructed the basic theorems of single-copy LOCC and SLOCC
transformation, Neilsen's and Vidal's theorem in the
infinite-dimensional pure state space (Theorems \ref{epsilon
Nielsen} and~\ref{epsilon Vidal}). As a result we showed that
under this change of definition the framework of entanglement
convertibility is preserved for infinite-dimensional systems, and
therefore, our definition of $\epsilon$-convertibility for LOCC is
suitable and sufficient for realistic conditions of quantum
information processing in infinite-dimensional systems.

In Section \ref{Extension} in order to study SLOCC convertibility
in infinite-dimensional systems, we constructed a pair of SLOCC
monotones which can be considered as extensions of the Schmidt
rank to infinite-dimensional spaces. By these monotones we showed
that states with polynomially-damped Schmidt coefficients belong
to a higher rank of entanglement class than other states in terms
of single-copy SLOCC convertibility.

In the last Section \ref{inhibition law} we showed that arbitrary
finitely many copies of exponentially-damped states cannot be
converted to even a single copy of polynomially-damped states.
Since such differences of classes do not exist in the
finite-dimensional setting, the SLOCC classification of
infinite-dimensional states has a much richer structure than for
finite-dimensional ones. Therefore, these new features of
entanglement have the potential to produce new quantum information
protocols which are impossible for finite-dimensional systems.
Finally, we stress that in infinite-dimensional systems, there
remain important problems that are yet to be solved even for the
simplest bipartite pure states.

\nonumsection{Acknowledgements} \noindent
MO is grateful to Professor M.\ Ozawa, Professor M.B.\ Plenio,
Professor K.\ Matsumoto, Professor M.\ Hayashi, and Dr.\ A.\
Miyake for discussions. This work has been supported by the Asashi
Grass Foundation, the Sumitomo Foundation, the Japan Society of
Promotion of Science, the Japan Scholarship Foundation, the
Japan Science and Technology Agency, 
and the Special Coordination Funds for Promoting Science and Technology.

\nonumsection{References} 

\appendix{\ Schmidt decomposition and Lo-Popescu's
  Theorem in infinite dimensional systems} \label{Lo-Popescu section}


In this appendix and the next, as a preparation for the proofs
of Nielsen's and Vidal's theorem in infinite dimensional systems,
we will see how we can extend basic theorems about LOCC and
majorization \cite{majorization, vidal} to infinite dimensional
systems.

At first, we extend the concept of Schmidt decomposition and
Schmidt coefficients in infinite-dimensional systems:
\begin{Theorem}{(Schmidt decomposition)}
For any $\ket{\Psi} \in \Hi = \Hi _A \otimes \Hi _B$, there exist
orthonormal sets ({\it but not necessarily  basis sets}) $\{
\ket{e_i} \} _{i=1}^{\infty} $ and $\{ \ket{f_i} \}
_{i=1}^{\infty} $ of $\Hi _A$, and $\Hi _B$, respectively, such
that
\begin{equation} \label{Schmidt inf}
     \ket{\Psi} = \sum _{i=1}^{\infty} \sqrt{\lambda _i} \ket{e_i}
     \otimes \ket{f_i} \:,
\end{equation}
where $\lambda _i \ge 0$, $\lambda _i \ge \lambda _{i+1} $ and
$\sum _{i=1}^{\infty} \lambda _i = 1$.  The representation of a
state $\ket{\Psi}$ in the form of Eq.(\ref{Schmidt inf}) is called
a Schmidt decomposition and $\{ \lambda _i \} _{i=1}^{\infty}$ are
called Schmidt coefficients in infinite-dimensional systems.
\end{Theorem}
\begin{Proof}
We use the singular value decomposition given as follows in an
infinite dimensional system: For a compact operator $M$ from $\Hi
_A$ onto $\Hi _B$, there exist orthonormal sets (but not necessarily
  basis sets) $\{ \ket{e_i} \}_{i=1}^{\infty} \subset \Hi _A$
and $\{ \ket{f_i} \}_{i=1}^{\infty} \subset \Hi _B$ and positive
real numbers $\{ \lambda _i  \}_{i=1}^{\infty}$ with
$\sqrt{\lambda _n} \rightarrow 0$ such that
\begin{equation}\label{Schmidt eq}
M = \sum _{i=1}^{\infty} \sqrt{\lambda _i}\ket{e _i} \bra{f _i},
\end{equation}
where the above sum converges in the operator norm \cite{functional
analysis}. In particular, if $M$ is a Hilbert-Schmidt class
operator, $\{ \sqrt{\lambda _i } \}_{i=1}^{\infty}$ satisfy $\sum
_{i=1}^{\infty} \lambda _i = (\| M \|_2)^2 \stackrel{\rm def}{=}
\Tr M^{\dagger}M$, where $\| \cdot \|_2$ is the Hilbert-Schmidt
norm \cite{functional analysis}. Thus, we derive Eq.(\ref{Schmidt
inf}) from Eq.(\ref{Schmidt eq}), because the linear map
$\ket{e_i} \bra{f_j} \mapsto \ket{e_i} \otimes \ket{f_j}$ gives an
isomorphism from the Hilbert-Schmidt space $\mathfrak{C} _2(\Hi
_A, \Hi _B)$ (the Hilbert space of all Hilbert-Schmidt class
operators between $\Hi_A$ and $\Hi _B$ with the inner product
$\left(M |N \right) \stackrel{\rm def}{=} \Tr M^{\dagger} N $) to
the Hilbert space $\Hi = \Hi _A \otimes \Hi _B$ \cite{functional
analysis}. \hfill $\square$
\end{Proof}

In finite $d$-dimensional bipartite systems, the Schmidt
decomposition of a state $\ket{\psi}$ is given by
\begin{equation} \label{Schmidt}
     \ket{\psi} = \sum _{i=1}^{d} \sqrt{\lambda _i} \ket{e_i}
     \otimes \ket{f_i}\;,
\end{equation}
where $\{ \ket{e_i} \} _{i=1}^{d}$ and $\{ \ket{f_i} \}
_{i=1}^{d}$ are the basis sets.  Therefore, convertibility of
states under local {\it unitary} operations are determined by the
Schmidt coefficients $\{ \lambda _i \} _{i=1}^{d}$ of states,
namely, the two states are convertible to each other under local
unitary operations if and only if the two states have  same
Schmidt coefficients.  In infinite-dimensional systems, the
Schmidt coefficients determine convertibility of states under
local {\it partial isometry} instead of local unitary operations.
That is, if the Schmidt coefficients of $ \ket{\Psi}$ and $\ket{\Phi}$
are the same, then there exist local partial isometries $U_A$ and $U_B$
and $\ket{\Psi} = U_A \otimes U_B \ket{\Phi}$ is satisfied.

Partial isometry is defined as a unitary operator between
subspaces. If we had defined the Schmidt coefficients to be a
sequence including the dimension of the kernel of the reduced
density matrix (of the given state), we could make Schmidt
coefficients indicating the convertibility under local unitary
operations. However, to develop the theory of LOCC and SLOCC, (which
include local partial isometries), convertibility for
infinite-dimensional systems, the former definition is more
suitable than the latter, so we take the definition of Eq.
{\ref{Schmidt inf}}. This is because states are convertible to
each other by LOCC, if and only if they are convertible to each
other by local partial isometries (we can show this fact from
Theorem \ref{infinite Nielsen}); moreover, there exists a pair of
states which are convertible to each other by local partial isometries,
but not by local unitaries. The following example satisfies such
a condition. Suppose states $\ket{\Psi}$ and $\ket{\Phi}$ on $\Hi _A
\otimes \Hi _B$ are defined as
\begin{eqnarray}
\ket{\Psi} &=& \sum _{i=1}^{\infty} \sqrt{\lambda _i} \ket{e_i}
\otimes \ket{f_i} \label{def of psi} \\
\ket{\Phi} &=& \sum _{i=1}^{\infty} \sqrt{\lambda _i} \ket{e_{2i}}
\otimes \ket{f_{2i}}, \label{def of phi}
\end{eqnarray}
where $\{ \ket{e_i } \}_{i=1}^{\infty}$ and $\{ \ket{f_i}
\}^{\infty}$ are orthonormal basis sets of $\Hi _A$ and $\Hi _B$,
respectively. In this case, $\ket{\Phi }$ and $\ket{\Psi }$ can be
convertible to each other by LOCC: $\ket{\Phi }$ can be
convertible to $\ket{\Psi }$ by a local partial isometry $\sum
_{i=1}^{\infty} \ket{e_{i}}\bra{e_{2i}} \otimes
\ket{f_i}\bra{f_{2i}}$, and $\ket{\Psi }$ can be convertible to
$\ket{\Phi }$ by a local isometry $\sum _{i=1}^{\infty}
\ket{e_{2i}}\bra{e_{i}} \otimes \ket{f_{2i}}\bra{f_{i}}$. However,
their are not convertible by local unitaries. This is because a
subspace spanned by $\{ \ket{e_{2i}}  \}_{i=1}^{\infty}$ and a
subspace spanned by $\{ \ket{f_{2i}}_{i=1}^{\infty}$ should be mapped to $\Hi
_A$ and $\Hi _B$, respectively; a proper subspace should be mapped
to a whole space. This map is obviously impossible by unitary
operators, which are bijections and always map a whole space to a
whole space.

Actually, from a physical point of view, convertibility under local
partial isometry can be understood that we may need additional
independent ancilla systems for each subspace to convert
$\ket{\Psi}$ to $\ket{\Phi}$. For example, $\ket{\Phi}$ defined as
Eq.(\ref{def of phi}) can be transformed to $\ket{\Psi}$ defined
as Eq.(\ref{def of psi}) by the following protocol: First, attach
one-qubit ancilla systems $\Hi _{A'}$ and $\Hi _{B'}$ to both
local systems $\Hi _A$ and $\Hi _B$, and prepare the ancilla
systems in $\ket{0}_{A'}$ and $\ket{0}_{B'}$. Second, apply  a
local unitary transformation $U_{AA'} \otimes U_{BB'}$ to the
states $\ket{\Phi}_{AB} \otimes \ket{0}_{A'} \otimes
\ket{0}_{B'}$, where $U_{AA'}$ and $U_{BB'}$ are unitary
transformations on $\Hi _A \otimes \Hi _{A'}$ and $\Hi _B \otimes
\Hi _{B'}$ defined as
\begin{eqnarray*}
  U_{AA'}  &\stackrel{\rm def}{=}& \sum _{n=1}^{\infty} \big(
|n \rangle_{A} |0 \rangle _{A'} {}_{A} \langle 2n | {}_{A'} \langle 0| + 
 | 2n+1 \rangle _A |1 \rangle _{A'} {}_{A}\langle 2n+1| {}_{A'}\langle 0|  \\
&\quad & \qquad  + |2n \rangle_A |1 \rangle _{A'} {}_A \langle n| 
{}_{A'}\langle 1| \big) \\
  U_{BB'} &\stackrel{\rm def}{=}& \sum _{n=1}^{\infty} \big(
|n\rangle_{B} |0\rangle_{B'} {}_B \langle 2n| {}_{B'}\langle 0| + 
| 2n+1 \rangle_B |1\rangle_{B'} {}_B\langle 2n+1| {}_{B'}\langle 0|  \\
&\quad & \qquad + |2n\rangle_B |1\rangle_{B'} {}_B\langle n| {}_{B'}\langle 1| \big).
\end{eqnarray*}
After this local unitary transformation, the state is changed to
$\ket{\Psi} _{AB} \otimes \ket{0}_{A'} \otimes \ket{0}_{B'}$.
Finally, by removing the ancilla system $\Hi _{A'}$ and $\Hi
_{B'}$, we derive $\ket{\Psi}$ on the systems $\Hi _A \otimes \Hi
_B$.

  For finite-dimensional systems, Lo and Popescu proved
that if $\ket{\Psi}$ can be converted into $\ket{\Phi}$ by LOCC,
there exists a one-way classical communication LOCC which consists
of a local measurement of one of the local spaces and a unitary
operation of the other depending on the result of the measurement
\cite{lo-popescu}. This theorem is called the Lo-Popescu theorem.
Intuitively speaking, the Schmidt decomposition denotes the existence
of symmetry between local subspaces, and Lo-Popescu's theorem is
the reflection of the symmetry of subsystems. As we have shown that
the Schmidt decomposition in infinite-dimensional systems is
weaker (indicating equivalence under partial isometries instead of
unitary operations) than finite-dimensional systems, the
corresponding Lo-Popescu theorem is slightly modified as
follows.

\begin{Theorem}[Lo-Popescu's] \label{Lo-Popescu}
In the separable Hilbert space $\Hi _A \otimes \Hi _B$ , for any
given state $\ket{\Psi}$  and bounded operator $M \in \B (\Hi _B)$
there exist a bounded operator $N \in \B (\Hi _A)$ and {\it partial
isometry} $U \in \B (\Hi _B)$ which satisfy $ I \otimes M
\ket{\Psi} = N \otimes U \ket{\Psi}$.
\end{Theorem}

\begin{Proof}
Suppose $\ket{\Psi} = \sum _{i=1}^{\infty} \sqrt{\mu _i} \ket{a_i}
\otimes \ket{b_i}$. Define a partial isometry $U$ as $U= \sum
_{i=1}^{\infty} \ket{b_i} \bra{a_i}$, then we have
\begin{equation}
I \otimes M \ket{\Psi} =\sum _{i=1}^{\infty} \sqrt{\mu _i}
\ket{a_i} \otimes  M \ket{b_i} \nonumber \;,
\end{equation}
and
\begin{equation}
MU\otimes I \ket{\Psi} =\sum _{i=1}^{\infty} \sqrt{\mu _i} M
\ket{b_i} \otimes  \ket{b_i}\nonumber \;.
\end{equation}
Thus, we obtain
\begin{eqnarray}
\Tr\; _A I \otimes M \ket{\Psi} \bra{\Psi} I\otimes M^{\dagger} =
\Tr\; _B MU \otimes I \ket{\Psi} \bra{\Psi} U^{\dagger}
M^{\dagger} \otimes I\;.\nonumber
\end{eqnarray}
  By our definition of Schmidt decomposition, $\rho _A$ and $\rho
_B$ are partial isometry equivalent for any state $\ket{\Psi}$.
Therefore, there exist partial isometries $U_A$ and $U_B$ which
satisfy
\begin{equation}
U_A \otimes U_B (MU \otimes I) \ket{\Psi} = I \otimes M
\ket{\Psi}\nonumber\;.
\end{equation}
Defining $N \stackrel{\rm def}{=} U_A MU \otimes U_B $, then the
theorem has been proven. \hfill $\square$
\end{Proof}
%


\appendix{ \quad Majorization in infinite-dimensional systems}
\label{Majorization}

In finite-dimensional systems, majorization is a pseudo partial
ordering on the whole vector space \cite{Bhartia}. On the other
hand in infinite-dimensional system, majorization can be defined
on only a subset of the whole vector space. To formulate our
definition of Schmidt coefficients in infinite-dimensional
systems, we define a majorization on
\begin{equation}
     l _1 = \{ \{ x_i \} _{i=1}^{\infty} | \sum _{i=1}^{\infty} |x_i| <
\infty \}\;.
\end{equation}
For mathematical simplicity, we only define majorization on
\begin{eqnarray}
     l_1 ^+ = \{ \{ x \} _{i=1}^{\infty } \in l_1 | x_i \ge 0,
  \sharp \{ i | x_i=0 \} < \infty \nonumber \vee \ \sharp \{ i | x_i > 0 \} < \infty \}\;,
\end{eqnarray}
where $\sharp$ denotes the cardinality of a set. In this case
$l_1^+$ is a convex cone of $l_1$ and identifies the set of all
permutations of Schmidt coefficients. 
Thus, $l_1^+$ is enough for our purposes, and we do not need 
to define majorization for all element of $l_1$. Moreover, 
in the following definition of majorization, we use decreasing reordering 
of $x \in l_1$. However, if $x$ is not in $l_1^+$, 
it is difficult to rearrange elements of
$x$ in decreasing order, 
and we need to extend the definition of majorization
so as to include sequences which are not in $l_1^+$, but in $l_1$.\fnm{{\it a}}\fnt{}{\ 
The general definition of majorization of sequences can be found in
\cite{markus}. The all propositions in this subsection can be extended to this general case 
by appropriate modification of proofs.} This is also the reason why we
define majorization only on $l_1^+$.

  Now, we define majorization
of Schmidt coefficients in infinite-dimensional systems as
follows:
\begin{Definition}
For any $x , y \in l_1^+$, $x \prec _{\omega} y $ (or $x$ is
sub-majorized by $y$) is defined, if and only if
\begin{equation}
     \sum _{i=1}^k x^{\downarrow} _i \le
     \sum _{i=1}^k y^{\downarrow}_i\;,
\end{equation}
for $k \in \mathbb{N}$, where $x^{\downarrow} _i$ is given by
$x^{\downarrow} _i = x_{P(i)}$, and $P$ is an element of an infinite
symmetry group satisfying $x^{\downarrow} _i \ge x^{\downarrow}
_{i+1}$ in decreasing reordering of $x$. Similarly, if $x$ and $y$
satisfy,
\begin{equation} \label{super-majorization}
\sum _{i=k}^{\infty} x^{\downarrow} _i \ge \sum _{i=k}^{\infty}
x^{\downarrow} _i\;,
\end{equation}
then, we write $x \prec ^{\omega} y$ and say $x$ is
super-majorized by $y$.

Additional to the sub-majorization or super-majorization conditions,
(both conditions lead to the same majorization condition) if $x$ and
$y$ satisfy the normalization condition
\begin{equation}
     \sum _{i=1}^{\infty} x^{\downarrow} _i
     = \sum_{i=1}^{\infty} y^{\downarrow} _i\;,
\end{equation}
then we write $x \prec y$ and say $x$ is majorized by $y$.
\end{Definition}
The sub-majorization condition does not require the normalization
condition, but it is easily proven that $\mathbf{x} \prec
_{\omega} \mathbf{y} $ is equivalent to
\begin{eqnarray}
\sum _{j=1}^{\infty} (\mathbf{x_j^{\downarrow} }  -t)^+ \le
  \sum _{j=1}^{\infty} (\mathbf{y_j^{\downarrow} } -t)^+\;,
\end{eqnarray}
for all real $t$, where $z^+ = \max (z,0) $ is the positive part of
any real number.

Uhlmann's theorem relates operations on quantum states and
majorization conditions.  This theorem is one of the essential items
for proving Nielsen's theorem for LOCC convertibility.   To prove
Uhlmann's theorem in infinite-dimensional systems,  we define a
doubly stochastic matrix in infinite-dimensional systems. It is
similar to the one for finite-dimensional systems.  For all double
sequences $\{ d_{ij} \}_{ij = 1}^{\infty}$ which satisfy
$\sum_{i=1}^{\infty} d_{ij} =1$ for all $j$, and $\sum
_{j=1}^{\infty} d_{ij} = 1$ for all $i$, we can define a bounded
linear operator $D \in \B (l_1)$ by
\begin{eqnarray}
     D \mathbf{x} = \{ \sum_{j=1}^{\infty} d_{ij}
     \mathbf{x_j} \} _{i=1}^{\infty}\;,
\end{eqnarray}
for all $\mathbf{x} \in l_1$. These operators are called doubly
stochastic matrices on $l_1$.  We can easily see that the operator
norm of a doubly stochastic matrix is $1$.

The defined doubly stochastic matrices are related to majorization
as follows: If $D$ is doubly stochastic, then, for all $\mathbf{x} \in
S$, we have
\begin{eqnarray} \label{ds}
     D \mathbf{x} \prec \mathbf{x}\;,
\end{eqnarray}
since
\begin{eqnarray}
     \sum_{i=1}^{\infty} [(Dx)_i - t ]^+ &=&
     \sum_{i=1}^{\infty}
     \Bigl[ \sum_{j=1}^{\infty} D_{ij} ( x_j - t ) \Bigr]^+ \nonumber\\
      &\le&  \sum_{i=1}^{\infty}
      \sum_{j=1}^{\infty} D_{ij} ( x_j - t )^+ \nonumber\\
      &=& \sum_{j=1}^{\infty} (x_j - t )^+\;,
\end{eqnarray}
for any real $t$, due to the convexity of $( )^+$.  On the other hand,
$\sum_{i=1}^{\infty} (Dx)_i = \sum_{i=1}^{\infty} x_i$ is trivial.
Thus, the necessary condition (majorization condition) for the doubly
stochastic condition is proven.  We note that Eq.(\ref{ds}) is
also valid for weaker conditions of $D$ than the doubly stochastic
matrix, for example, $D$ satisfying $\sum _{j=1}^{\infty} d_{ij}
\le 1$.

Now we are ready to extend Uhlmann's theorem for
infinite-dimensional systems:
\begin{Theorem}[Uhlmann] \label{Uhlmann}
If the two density operators $\rho _1$ and $\rho _2$ on the infinite
separable Hilbert space $\Hi$ satisfy the following relation,
\begin{eqnarray}
\rho _1 = \sum _{j=1}^{\infty} p_j U_j \rho _2 U_j^{\dagger}, \label{Uhlmann eq}\\
\qquad \sum _{j=1}^{\infty} p_j = 1 ,
\end{eqnarray}
where $U_j$ are partial isometries whose initial space includes
closure of the range of $\rho _2$, ${\rm ker U_j}^{\perp} \supset
\overline{\rm Ran \rho _2} $, then the non-zero eigenvalue of $\rho
_1$ is majorized by $\rho _2$.
\end{Theorem}

\begin{Proof}
Suppose $\rho _1 = \sum _{k=1}^{\infty} \mu _k \ket{e_k}
\bra{e_k}$ and $\rho _2 = \sum _{i=1}^{\infty} \lambda _i  \ket{i}
\bra{i}$. Define a partial isometry as $V = \sum _{i=1}^{\infty}
\ket{i} \bra{e_i}$, whose initial space is given by $\overline{\rm
Ran \rho _1}$, and whose final space is given by $\overline{\rm Ran
\rho _2}$. Then $U_j V$ is a partial isometry whose initial and
final space are given by $\overline{\rm Ran \rho _1}$ and $U_j(
\overline{\rm Ran \rho _2} )$, respectively. Actually, it is
trivial that $U_j V$ is a zero operator on ${ \overline{\rm Ran
\rho} }^{\perp}$. Now suppose $\ket{\Phi} \in \overline{\rm Ran
\rho _1}$, then we obtain $\| U_j V \ket{\Phi} \| = \| \ket{\Phi}
\|$ from the condition $V \ket{\Phi} \in \overline{\rm Ran \rho
_2}$.

Next, define Fourier's coefficients of $U_j \ket{i}$ as $u
_{ji}^h$, that is, $U_j \ket{i} = U_j V \ket{e_i} =\sum
_{h=1}^{\infty} u _{ji}^h \ket{e_h}$, and
\begin{equation} \label{normalization of u}
  \sum _{h=1}^{\infty} | u _{ji}^h | ^2 =1.
\end{equation}
Then, we can rewrite Eq.(\ref{Uhlmann eq}) as,
\begin{equation}
\sum _{k=1}^{\infty} \mu _k \ket{e_k} \bra{e_k} = \sum _{ij \in
N^2} p_j \lambda _i (\sum _{h=1}^{\infty} u _{ji}^h \ket{e_h} )
(\sum _{l=1}^{\infty} {u^{\ast}} _{ji}^l \bra{e_l} )\;.
\end{equation}
Taking the inner product between $\ket{e_n}$,
\begin{equation}
     \mu _n = \sum _{ij \in \mathbb{N}^2} p_j \lambda _i | u _{ji}^n | ^2
        = \sum _{i=1}^{\infty} \lambda _i \sum _{j=1}^{\infty} p_j
        | u_{ji}^n | ^2\;.
\end{equation}
We define $D _{ni}$ as $D _{ni} = \sum _{j=1}^{\infty} p_j |
u_{ji}^n | ^2$. (\ref{normalization of u}) guarantees that $\sum
_{n=1}^{\infty} D _{ni} = 1$. Since
\begin{eqnarray}
\sum _{i=1}^{\infty} | u_{ji}^n |^2 &=& \sum _{i=1}^{\infty}
| \bra{e_n} U_j V \ket{e_i} |^2 \le  \| U_j V \ket{e_n} \| ^2 \nonumber\\
  &\le& \| U_j \|_{\rm op} ^2 \| V \|_{\rm op} ^2 \| \ket{e_n} \| ^2 =1\; ,
\end{eqnarray}
where $\| \cdot \|_{\rm op}$ is the operator norm and $\sum
_{i=1}^{\infty} D _{ni} \le 1$.
  Therefore, using the
necessary condition of the double stochastic matrices and the
weaker condition of (\ref{ds}) for $D$, we derive $\mu = D \lambda
\prec \lambda$. The theorem is proven.
\end{Proof}

\newpage
\appendix{ \quad Proof of Theorem \ref{epsilon Nielsen} (Nielsen's theorem
in infinite-dimensional systems)} \label{Proof of Nielsen}

In this appendix, based on Appendix A and B, we prove Nielsen's
\cite{majorization} theorem for infinite dimensional systems.

Before we show the proof of Nielsen's theorem of
$\epsilon$-convertibility in infinite-dimensional systems, we
first show that the necessary part of the original proof
of Nielsen's theorem \cite{majorization} can be directly extended
to infinite-dimensional systems by means of the Lo-Popescu and
Uhlmann theorems we have already proven in the last section.
\begin{Lemma} \label{necessity Nielsen}
The necessary condition for convertibility of an
infinite-dimensional state $\ket{\Psi}$ to another state
$\ket{\Phi}$ under LOCC operations is given by $\lambda \prec
\mu$, where $\lambda = \{ \lambda \} _{i=1}^{\infty}$ and $\mu =
\{ \mu \} _{i=1}^{\infty} $ are the sequences of Schmidt
coefficients of the states $\ket{\Psi}$ and $\ket{\Phi}$,
respectively.
\end{Lemma}
\begin{Proof}
Suppose $\ket{\Psi}$ can be converted to $\ket{\Phi}$ by LOCC,
then By Lo-Popescu's theorem, $\rho _{\Phi} = p_m M_m \rho _{\Psi}
M _m^{\dagger}$ where $\sum _{m=1}^{\infty } p_m = 1$, $\rho
_{\Psi} = \Tr\; _B ( \ket{\Psi} \bra{\Psi} )$ and $\rho _{\Phi}
=\Tr\; _B ( \ket{\Phi} \bra{\Phi} )$. Then, according to the same
method of Nielsen's original proof, we derive
  $\rho _{\Psi} = \sum
_{m=1}^{\infty } p_m U_m \rho _{\Phi } U_m^{\dagger }$, where
$U_m$ is a partial isometry originating in the polar decomposition
of $M_m \sqrt{\rho _{\Psi }}$. Since $\ker U_m ^{\perp } = \ker
M_m \sqrt{\rho _{\Psi }}^{\perp } \supset \overline{\rm Ran \rho
_{\psi } }$ \cite{functional analysis}, by Uhlmann's theorem we
get $\lambda \prec \mu$. \hfill $\square$
\end{Proof}

By means of this lemma, we can prove the necessary part
of Nielsen's theorem in infinite-dimensional systems. For the
sufficient condition, we fully use the properties of
$\epsilon$-convertibility.
\begin{Proof}{(Theorem \ref{epsilon Nielsen})}

Only if part: If $\ket{\Psi}$ is $\epsilon$-convertible to
$\ket{\Phi}$ for any $\epsilon > 0$, there exists a sequence of
states $\{ \ket{\Phi' _n} \} _{n=1}^{\infty}$ which strongly
converges to $\ket{\Phi}$ (for pure states the topology of the trace norm
is stronger than the strong topology of Hilbert space). Then, from
Lemma \ref{necessity Nielsen}, $\lambda \prec \mu ^{'} _n$ where
$\mu ^{'} _n$ and $\lambda$ are the Schmidt coefficients of
$\ket{\Phi' _n}$ and $\ket{\Psi}$ for all $n \in \mathbb{N}$.
Because Schmidt coefficients are continuous in the strong topology,
$\sum _{i=1}^{n} \mu ^{'} _{n,i} \ge \sum _{i=1}^{n} \lambda  _i$
means $\sum _{i=1}^{n} \mu  _i \ge \sum _{i=1}^{n} \lambda  _i$
where $\mu _i$ are the Schmidt coefficients of $\ket{\Phi}$.

If part: When the Schmidt ranks (the number of non-zero Schmidt
coefficients) of both $\ket{\Psi}$ and $\ket{\Phi}$ are finite,
the proof is identical to the one for finite-dimensional systems.
By the definition of Schmidt decomposition, we can assume
$\ket{\Psi}$ and $\ket{\Phi}$ have the same Schmidt basis without
loss of generality.  In what follows, we divide the proof for
the remaining cases into two parts: the case
where $\ket{\Psi}$ has finite Schmidt rank, and
the case where both of the states have infinite Schmidt ranks.

1)  The case that $\ket{\Psi}$ has infinite Schmidt rank and
$\ket{\Phi}$ has finite Schmidt rank:\\
Suppose the Schmidt rank of $\ket{\Phi}$ is given by $N$. In what
follows, we assume $\epsilon$ is arbitrary, but satisfies
$\epsilon < \mu _N$. Since for any Schmidt coefficient $\{ \lambda
_i \}_{i=1}^{\infty}$, we have $\lim _{n \rightarrow \infty} n
\lambda _n = 0$. Therefore, there exists an $N_1 (\epsilon ) $ such
that $n \lambda _n < \epsilon /2$ for any $ n \ge N_1$. On the
other hand, since $\sum _{i=1}^{\infty} \lambda _i =1$ , there
exists an $N_2 (\epsilon )$ such that $\sum _{i=n}^{\infty} \lambda
_i < \epsilon /2$ for any $ n \ge N_1$. Suppose $M = \max (N_1,
N_2, N)$, then we define $\{ {\mu '}_i \}$ as follows:
\begin{tabbing}
\qquad \= For \= $1 \le i \le N-1 $ \quad \= : ${\mu '}_i = \mu _i$ \\
\>   \> $i=N$              \> : ${\mu '}_N = \mu _N - \left ((M-N)
     \lambda _M +  \sum _{n=M+1}^{\infty} \lambda _n \right)$ \\
\>   \> $N+1 \le i \le M$  \> : ${\mu '}_i = \lambda _M$ \\
\>   \> $M+1 \le i $       \> : ${\mu ' }_i = \lambda _i$
\end{tabbing}
We define $\ket{\Phi '}$ as $\ket{\Phi '} = \sum _{i=1}^{\infty}
\sqrt{{\mu '}_i} \ket{i} \otimes \ket{i}$. Then, by definition,
$\lambda \prec \mu^{'} \prec  \mu $. Moreover, we
obtain
\begin{eqnarray}
\| \ket{\Phi} - \ket{\Phi '} \| ^2 &=& |(M-N)\lambda _M + \sum
_{n=M+1}^{\infty} n \lambda _n | ^2  \nonumber \\
&=& |(M-N)\lambda _M|^2
+ |\sum _{i=M+1}^{\infty} \lambda_i|^2 \nonumber \\
&\le & \epsilon ^2\;.
\end{eqnarray}
Therefore, for any neighborhood of $\ket{\Phi}$, we can find
a $\ket{\Phi'}$ such that $\ket{\Psi} \rightarrow \ket{\Phi'}$.

2) The case that the Schmidt ranks of $\ket{\Psi}$
and $\ket{\Phi}$ are infinity:\\
By easy calculation we can show that for any $\epsilon$, there
exists a natural number $N_1 (\epsilon )$ such that if $\ket{\Phi
'} = \sum _{i=1}^{\infty} \sqrt{{\mu '} _i} \ket{i} \otimes
\ket{i}$ satisfies $\mu ' _i= \mu$, $i \le N_1 (\epsilon ) $,
then $\| \ket{\Phi}\bra{\Phi} - \ket{\Phi '}\bra{\Phi'} \|_{\rm
tr} < \epsilon $. Since $\sum _{i=1}^n \lambda _i < 1$ and $\lim
_{n \rightarrow \infty} n \lambda _n = 0$ for any $n  \in
\mathbb{N}$,
\begin{equation}
   \lim _{N_2 \rightarrow \infty} \Bigl[\sum _{k=1}^{N_2 (\epsilon )} \lambda _k
   - (N_2 - N_1 )\lambda _{N_2} \Bigr] = 1\;.
\end{equation}
Thus, there exists a natural number $N_2 (\epsilon ) \ge N_1
(\epsilon ) + 1 $ such that
\begin{equation} \label{eq N_2}
\sum _{k=1}^{N_2 (\epsilon )} \lambda _k
   - (N_2 - N _1 ) \lambda _{\lambda _{N_2}}
   \ge \sum _{i=1}^{N_1 (\epsilon ) } \mu _i\;,
\end{equation}
and
\begin{equation} \label{epsilon 1}
\sum _{k=1}^{N_2 (\epsilon ) - 1} \lambda _k
  - (N_2 - N _1 -1) \lambda _{N_2 - 1}
   \le \sum _{i=1}^{N_1 (\epsilon ) } \mu _i\;.
\end{equation}
We examine in the two cases $N_2(\epsilon ) = N_1(\epsilon ) +1$
and $N_2 (\epsilon ) > N_1(\epsilon ) + 1$ separately.

a)  The case that $N_2(\epsilon ) = N_1(\epsilon ) +1$:  The
inequalities (\ref{eq N_2}) and (\ref{epsilon 1}) guarantee $\sum
_{k=1}^{N_1 (\epsilon )} \lambda _k = \sum _{k=1}^{N_1(\epsilon )}
\mu _k$. If we define $1 \le k \le N_1(\epsilon )$, ${\mu '}_k =
\mu _k$, $k \ge N_1(\epsilon ) + 1$ and  ${\mu '}_k = \lambda _k$,
$\{ {\mu '}_i \}_{i=1}^{\infty}$ satisfies $\sum _{i=1}^k \lambda
_i \le \sum _{i=1}^k {\mu '}_i \le \sum _{i=1}^k \mu _i$ for all
$k \in \mathbb{N}$.

b) The case that $N_2 (\epsilon ) > N_1(\epsilon ) + 1$: We define
$\delta$ as
\begin{eqnarray}
\delta = \Big[ \sum _{i=1}^{N_1 (\epsilon )} \lambda _i
   + \sum _{k=N_1 (\epsilon) +1}^{N_2 -1} (\lambda _k - \lambda _{N_2 })
  -\sum
_{i=1}^{N_1 (\epsilon ) } \mu _i \Big] /( N_2(\epsilon ) -
N_1(\epsilon ) -1) \:.
\end{eqnarray}
Then, since $\delta \ge 0$, we can define ${\mu '}_k$ as the
following,
\begin{tabbing}
\qquad \= ${\mu '}_k = \mu _k$ \qquad \qquad \= for $1 \le k \le N_1 (\epsilon )$, \\
  \> ${\mu '}_k = \lambda _{N_2(\epsilon )} + \delta$ \> for $N_1(\epsilon )+1 \le k \le N_2(\epsilon ) -1$, \\
  \> ${\mu '}_k = \lambda _k$  \> for $N_2(\epsilon ) \le k$
\end{tabbing}
then we have
\begin{eqnarray}
     \sum _{i=1}^{\infty} {\mu '}_i &=&
     \sum _{i=1}^{N_1 (\epsilon )}
     \mu _i + \sum _{k=N_1 (\epsilon )+1}^{N_2 (\epsilon )-1}
     (\lambda_{N_2(\epsilon )} + \delta )
     +\sum _{k=N_2(\epsilon )}^{\infty}
     \lambda _k \nonumber\\
&=&1\;.
\end{eqnarray}
Therefore, the $\{ \mu '_i \}_{i=1}^{\infty}$ are well defined  Schmidt
coefficients.

First, we show $\sum _{i=1}^{N} \mu '_i \le \sum _{i=1}^{N} \mu
_i$ for all $N $. Since this condition is trivial for $N \le N_1$
and $N_2 \le N$ by definition of $\{ \mu' _i \}_{i=0}^{\infty}$,
we only need to check this condition for $ N_1 + 1 \le N \le N_2 -
1$. Suppose there exists an $N'$ such that $N_1 +1 \le N' \le N_2 -1$
and $\sum _{i= N_1 +1}^{N'} \mu'_i > \sum _{i = N_1 +1}^{N'}\mu
_i$. Then, since $\mu'_i = \lambda _{N_2} + \delta$ for all $N_1
+1 \le i \le N_2 - 1$ and $\mu_i \ge \mu_{i+1} $, we can easily
see $\mu' _{N'}
> \mu _{N'}$. That is, $\sum _{i= N_1 +1}^{N} \mu'_i > \sum _{i = N_1 +1}^{N}\mu
_i$ for all $N' \le N \le N _2 -1$, and we can conclude $\sum _{i
= N_1 +1}^{N_2 -1} \mu'_i > \sum _{i=N_1}^{N_2 -1} \mu _i$ which
is a contradiction. Therefore, for all $N_1 + 1 \le N  \le N_2 -1 $,
$\sum _{i = N_1 +1}^{N} \mu'_i \le \sum _{i=N_1}^{N} \mu _i$.

Second, we show $\sum _{i=1}^N \lambda _i \le \sum _{i=1}^N \mu'
_i$ for all $N$. Since this condition is trivial for $N \le N_1$
and $N_2 \le N$, we only check for $N_1 +1 \le N \le N_2 -1$. In
this case, we calculate
\begin{eqnarray}
\sum _{k=1}^N {\mu '}_k - \sum _{k=1}^N \lambda _k &=& \sum
_{k=1}^{N_1} \mu _k + \sum _{k=N_1+1}^N (\lambda _{N_2} + \delta)
- \sum _{k=1}^N \lambda _k \nonumber \\  &=& \frac{N_2 - N -
1}{N_2 - N_1 - 1} [(\sum _{k=1}^{N_1} \mu _k - \sum _{k=1}^{N_1}
\lambda _k ) - \sum _{k=N_1 +1}^N \lambda _k ] \nonumber \\
&\ & + \frac{N-N_1}{N_2 - N_1 -1} \sum _{k=N+1}^{N_2-1} \lambda
_k\;. \label{epsilon 3}
\end{eqnarray}
 From Eq.~(\ref{epsilon 1}), we obtain
\begin{equation}
     \sum _{k=1}^{N_1} \lambda _k + \sum _{k=N_1 + 1}^{N_2 -2} (\lambda _k
     - \lambda _{N_2} ) \le
      \sum _{k=1}^{N_1} \mu _k\;.
\end{equation}
Thus, for the Eq.(\ref{epsilon 3}) may be bounded by
\begin{eqnarray*}
  & \quad & \sum _{k=1}^N \mu' _k - \sum _{k=1}^N \lambda _k \\
&\ge &\frac{N_2 - N -1 }{N_2 - N_1 -1} \left[ \sum _{k= N_1
+1}^{N_2 -2} (\lambda _k - \lambda _{N_2 -1}) - \sum _{k= N_1 +
1}^{N} \lambda _k \right] + \frac{N - N_1}{N_2 - N_1 -1} \sum _{k=
N+1}^{N_2 -1}
\lambda _k. \\
&=& \sum _{k= N+1}^{N_2 -2} \lambda _k - (N_2 -N  -2 )\lambda
_{N_2 -1} \\
&\ge& 0.
\end{eqnarray*}
Thus, for $N_1 +1 \le N \le N_2 -1$, we have $\sum _{k=1}^N \lambda _k
\le
  \sum _{k=1}^N {\mu '}_k$.

Finally, for any natural number $N$, we obtain
\begin{equation}\label{}
     \sum _{k=1}^N \lambda _k \le \sum _{k=1}^N {\mu '}_k \le \sum
_{k=1}^N \mu _k\;.
\end{equation}
We define $\ket{\Phi'}$ by using Schmidt coefficients $\{ \mu' _i
\}_{i=1}^{\infty}$. Then, since $\mu'_k = \lambda _k$ for all $k
\ge N_2(\epsilon )$, we can convert $\ket{\Psi } $ to
$\ket{\Phi'}$ by means of the LOCC operation which is derived by
the original (finite dimensional) Nielsen's
theorem \cite{majorization}. Moreover, since $\mu' _k = \mu _k$ for
all $k \le N_1(\epsilon )$, $\ket{\Phi'}$ satisfies $\|
\ket{\Phi'}\bra{\Phi'} - \ket{\Phi}\bra{\Phi} \|_{\rm tr} <
\epsilon$. Therefore, for any neighborhood of $\ket{\Phi}$, we can
find a $\ket{\Phi'}$ such that $\ket{\Psi} \rightarrow \ket{\Phi'}$,
where $\ket{\Phi'}$ is defined by using Schmidt coefficients $\{
\mu' _i \}_{i=1}^{\infty}$.

\hfill $\square$
\end{Proof}

\appendix{ \quad Proof of Theorem \ref{epsilon Vidal}} \label{Proof of Vidal}
In this appendix, based on Appendix A, B and C, we prove Vidal's
\cite{vidal} theorem infinite dimensional systems.

\begin{Proof}
If part: The proof of Vidal's theorem in \cite{D'ariano} is most
suitable to extend this part to infinite-dimensional systems.
Suppose $\lambda \prec ^{\omega} p \mu$, then $\{ \nu _i \}
_{i=1}^{\infty}$ defined by the condition $\nu _1 = 1-p(1-\mu_1)$
and $\nu _i = p \mu _i$ for $i \neq 1$ satisfies the conditions
$\lambda \prec \nu$ and $\nu \le \mu$. If we define the state
$\ket{\Omega}$ as the state whose Schmidt coefficients are $\{
\nu _i \}$ and whose Schmidt basis is same as $\ket{\Phi}$'s,
then, by Nielsen's theorem in infinite-dimensional systems, for
any small $\epsilon >0$, $\ket{\Psi}$ can be transformed to
$\ket{\Omega'}$ by LOCC with certainty, where $\|
\ket{\Omega'}\bra{\Omega'} - \ket{\Omega}\bra{\Omega} \| \le
\epsilon $.  Then, $\ket{\Omega'}$ can also be transformed to
$\ket{\Phi'}$  by the local measurement $E = \sum _{i=1}^{\infty}
\sqrt{\frac{\nu _i}{\mu _i}} \ket{i} \bra{i}$ with probability
$p$, where $\| \ket{\Phi'}\bra{\Phi'} - \ket{\Phi}\bra{\Phi} \|
_{\rm tr} \le \epsilon $.

Only If part: At first, since the first half of theorem 2 of
\cite{vidal} is directly extended to infinite-dimensional systems,
Vidal's monotone $E_n(\ket{\Psi}) = \sum _{i=n}^{\infty} \lambda
_i$ is a monotonic function of LOCC in mean value. In other words,
if $\ket{\Psi}$ can be transformed to $\ket{\Phi _i}$ with
probability $p_i$ in LOCC, then $E_n(\ket{\Psi}) \ge \sum
_{i=1}^{\infty} p_i E_n(\ket{\Phi _i})$.

If for a set of $\ket{\Psi}$ and $\ket{\Phi}$, and any $\epsilon
_1 > 0$, there exists a $\ket{\Phi'}$ such that $\|
  \ket{\Phi'} \bra{\Phi'} - \ket{\Phi} \bra{\Phi} \|_{\rm tr} <
\epsilon _1$ and $\ket{\Psi}$ can be converted to $\ket{\Phi'}$ by
some SLOCC with probability $p'$ which satisfies $p' \ge p$ and
$\lambda \prec ^{\omega} p\mu$ is not true, that is for some $k_1$
$E_{k_1} (\ket{\Psi}) < p E_{k_1} (\ket{\Phi})$. Then, there exists
a sequence $\ket{\Phi _n}$ and $p_n$ which satisfies  $\lim _{n
\rightarrow \infty} \ket{\Phi _n } = \ket{\Phi}$ and $p_n \ge p$ for all $n
\in \mathbb{N}$. Moreover, there also exists a sequence of SLOCC
operations which transforms $\ket{\Psi}$ to $\ket{\Phi _n}$. Then,
by monotonicity of $E_k (\ket{\Psi})$ we have $E_k (\ket{\Psi})
\ge p_n E_k (\ket{\Phi _n})$ for all $k \in \mathbb{N}$. Since $1
- E_n (\ket{\Psi})$ is finite sum of eigenvalues of the reduced
density operator, $E_n(\ket{\Psi})$ is continuous in the whole
Hilbert space and for all $n$. Thus, by taking the limit of the
inequality, we have $E_k (\ket{\Psi}) \ge p E_k (\ket{\Phi})$ for
all $k \in \mathbb{N}$. This is a contradiction. \hfill $\square$
\end{Proof}

\end{document}